\documentclass[onecolumn]{revtex4}
\usepackage{amsmath}
\usepackage{graphicx}

\begin{document}

\title{Scalar-tensor extension of Natural Inflation}

	\author{Guillem Simeon}
           \email{guillem.simeon@gmail.com}

	\affiliation{Facultat de F\'isica, Universitat de Barcelona, Mart\'i i Franqu\`es 1, 08028 Barcelona, Spain}

\begin{abstract}
We show that embedding Natural Inflation in a more general scalar-tensor theory, with non-minimal couplings to the Ricci scalar and the kinetic term, alleviates the current tension of Natural Inflation with observational data. The coupling functions respect the periodicity of the potential and the characteristic shift symmetry $\phi \rightarrow \phi + 2\pi f$ of the original Natural Inflation model, and vanish at the minimum of the potential. Furthermore, showing that the theory exhibits a rescaling symmetry in the regime where the coupling to the Ricci scalar is small and the coupling to the kinetic term is large, we obtain that the agreement with cosmological data can take place at an arbitrarily low periodicity scale $f$, solving at tree-level the problem of super-Planckian periodicity scales needed in Natural Inflation. 
\end{abstract}

\maketitle

\section{Introduction}

Natural Inflation [1] is a model for inflation protected against radiative corrections, based on a pseudo-Nambu-Goldstone inflaton, with a potential of the form $V(\phi)=\Lambda^{4}(1+\cos{(\phi/f)})$. Successful inflation is achieved by demanding the periodicity scale $f$ to take super-Planckian values and usually it is considered that $\Lambda \sim 10^{16}$ GeV, that is, $\Lambda$ is taken to be of the order of the Grand Unification Theories (GUT) scale.

A super-Planckian periodicity scale $f$ might be problematic due to corrections coming from gravitational instantons [2-5]. Besides this, in the light of the Planck collaboration results, standard Natural Inflation is in tension with cosmological data [6]. Some proposals have been made in order to tackle one or both issues (see [7-11], for example). 

Furthermore, in the context of inflationary cosmology, modifications of the gravitational sector have been considered in order to extend the single-field scenario and make it compatible with observations. These modifications give rise to a plethora of models. One special proposal, relevant for our present work, is to introduce a coupling of the inflaton to the Ricci scalar, thought to arise from quantum gravitational corrections (see [12] and references therein). One remarkable example of this class of theories is the Higgs inflation proposed by Bezrukov and Shaposhnikov [13].  These models are a subclass of the so-called scalar-tensor theories, where different couplings of curvature and a scalar field, the inflaton, are considered. We are also interested in the case where a derivative coupling to the Einstein tensor is introduced. These subclasses of theories have been studied in a cosmological setting ([14], [15], [16], [17], [18], [19]), and are known to give rise to accelerated expansion and equations of motion for the metric tensor and the scalar field with derivatives of order not higher than two.

In the framework of Natural Inflation, as we have mentioned before, some proposals have been put forward in order to solve the problems of the original model. One particular proposal, in the line of considering a coupling to the Ricci scalar, was to consider this coupling to be dynamic and to evolve with the same periodicity of the potential, thus preserving at tree-level the shift symmetry of Natural Inflation [20]. In the Einstein frame and in some specific cases $f \sim M_{P}$ and a better fit to Planck data could be obtained. Recently, the covariant one-loop quantum gravitational corrections to the effective potential of the model have been studied in [21], being a first step to roughly understand the implications of the presence of this periodic coupling to the Ricci scalar coming from the gravitational UV regime. On the other hand, the authors in [22] considered a non-canonical kinetic term with a coupling to the Einstein tensor, where $f \ll M_{P}$ could be obtained. In this case, the model is UV-protected, in the sense that all involved scales during inflation are much lower than the Planck scale. 

In this letter we consider the extension of Natural Inflation in the context of a scalar-tensor theory, where dynamic couplings to the Ricci scalar and to the  Einstein-tensor-coupled kinetic term are introduced. This proposal could be regarded as the minimal combination of [20] and [22], exploring the synergies between the two models, and motivated by the fact that from the EFT point of view shift-symmetric couplings should be considered in the action. The couplings are defined in such a way that respect the periodicity and the shift symmetry of standard Natural Inflation, and furthermore vanish at the minimum of the potential. We will show that this extension allows the scale $f$ to have arbitrarily low values and to obtain results in agreement with observational data from Planck, therefore solving at tree-level both of the aforementioned issues at the same time.

\section{Set-up}

In [23] the authors studied a general scalar-tensor theory and performed an analysis on the scalar and tensor perturbations of the model. In this letter, the theory will be restricted to the tree-level action:
\begin{equation}
S=\int d^{4}x \sqrt{-g} \bigg[\frac{1}{2}F(\phi)R -\frac{1}{2} {g}^{\mu\nu} \partial_{\mu}\phi \partial_{\nu}\phi+ {F}_{1}(\phi) G^{\mu\nu}  \partial_{\mu}\phi \partial_{\nu}\phi -V(\phi) \bigg],
\end{equation}

where ${G}^{\mu\nu} = {R}^{\mu\nu} - \frac{1}{2}{g}^{\mu\nu}R$ is the Einstein tensor and $V(\phi)={\Lambda}^{4}(1 + \cos{\frac{\phi}{f}})$. We have not considered the coupling to the Gauss-Bonnet invariant also included in [23], since we are interested in the minimal combination of the models [20] and [22]. We set $M_{P} = 1$ and we take the couplings to the Ricci scalar $R$ and the kinetic term to be, respectively,
\begin{equation}
 F(\phi) = 1 + \alpha\bigg(1+\cos{\frac{\phi}{f}}\bigg) = 1 + \frac{\alpha}{{\Lambda}^{4}}V
\end{equation}
\begin{equation}
 F_{1}(\phi) = \frac{\beta}{{\Lambda}^{4}}\bigg(1+\cos{\frac{\phi}{f}}\bigg) = \frac{\beta}{{\Lambda}^{8}}V,
\end{equation}

with two a priori arbitrary parameters $\alpha$ and $\beta$ measuring the strength of the couplings being introduced. Another reason for the exclusion of the Gauss-Bonnet coupling is to have more parametric control over the model, since there are already three free parameters ($\alpha$, $\beta$ and $f$), and we therefore leave the study of the Gauss-Bonnet coupling for future work. The choice of these coupling functions allows the theory described by (1) to conserve the characteristic shift symmetry $\phi \rightarrow \phi + 2\pi f$ of Natural Inflation. Furthermore, these couplings vanish at the minimum of the potential $\phi = \pi f$, reducing the theory to Einstein's gravity with a minimally coupled scalar field.

\vspace{3mm}

Variation of the action (1) with respect to the metric ${g}^{\mu\nu}$ and $\phi$ and considering a flat FRLW metric given by ${ds}^{2} = -{dt}^{2}+ {a(t)}^{2}({dx}^{2}+{dy}^{2}+{dz}^{2})$ gives the equations of motion of the model, describing an expanding homogeneous and isotropic universe. Defining the following slow-roll parameters

\begin{equation}
{\epsilon}_{0}=-\frac{\dot{H}}{H^{2}}, \qquad {\epsilon}_{1}=\frac{\dot{\epsilon_{0}}}{H {\epsilon}_{0}}
\end{equation}
\begin{equation}
{l}_{0}=\frac{\dot{F}}{HF}, \qquad {l}_{1}=\frac{\dot{l_{0}}}{H {l}_{0}}
\end{equation}
 
where dots denote differentiation with respect to time and $H \equiv \dot{a}/a$ is the Hubble parameter, and demanding $|{\epsilon}_{i}|,|{l}_{i}| \ll 1$, the equations of motion in the slow-roll regime are obtained [23]:  
 
\begin{equation}
3{H}^{2}F \approx V
\end{equation}
\begin{equation}
2\dot{H}F-H\dot{F} \approx -{\dot{\phi}}^{2} - 6{H}^{2}{F}_{1}{\dot{\phi}}^{2} 
\end{equation}
\begin{equation}
3H\dot{\phi} + V'- 6{H}^{2}F' + 18{H}^{3}{F}_{1}\dot{\phi} \approx 0
\end{equation}

where primes denote derivatives with respect to $\phi$. Equations (6) and (8) can be used to express the slow-roll parameters (4) and (5) as follows:
\begin{equation}
{\epsilon}_{0} \approx \frac{{V'}^{2}}{2{V}^{2}}\frac{(1-\tilde{\alpha}V)}{(1+\tilde{\alpha}V+2\tilde{\beta}{V}^{2})}, \quad {\epsilon}_{1}=\frac{\epsilon_{0}'}{{\epsilon}_{0}}\frac{\dot{\phi}}{H}
\end{equation}
\begin{equation}
{l}_{0}=\frac{F'}{F}\frac{\dot{\phi}}{H}, \qquad {l}_{1} = \frac{{l}_{0}'}{{l}_{0}}\frac{\dot{\phi}}{H}
\end{equation}

with $\tilde{\alpha} \equiv \alpha / {\Lambda}^{4}$, $\tilde{\beta} \equiv \beta/{\Lambda}^{8}$ and

\begin{equation}
\frac{\dot{\phi}}{H} \approx -\frac{V'}{V}\frac{(1+\tilde{\alpha}V)(1-\tilde{\alpha}V)}{(1+\tilde{\alpha}V+2\tilde{\beta}V^{2})}
\end{equation}
It is important to notice that all the quantities defined in (9), (10) and (11) are independent of the scale $\Lambda$.

\vspace{3mm}

According to the analysis of the scalar and tensor perturbations done in [23] in the general framework of a scalar-tensor theory, one can write the spectral tilt $n_{s}$ and the tensor-to-scalar ratio $r$ in our model as
\begin{equation}
n_{s} \approx 1-2{\epsilon}_{0}-{l}_{0} - \frac{2{\epsilon}_{0}{\epsilon}_{1} + {l}_{0}{l}_{1}}{2{\epsilon}_{0} + {l}_{0}}
\end{equation}
\begin{equation}
r \approx 8(2{\epsilon}_{0}+{l}_{0}).
\end{equation}

Slow-roll inflation ends whenever one of the aforementioned slow-roll parameters becomes $\simeq 1$, and we define the value of $\phi$ that fulfills this condition as ${\phi}_{e}$. Finding a value ${\phi}_{0}$ such that
\begin{equation}
N = \displaystyle\int_{{\phi}_{0}}^{{\phi}_{e}}{\frac{H}{\dot{\phi}} d\phi \approx 50 - 60},
\end{equation}
that is to say, finding the value ${\phi}_{0}$ that allows to drive $50-60$ e-folds of slow-roll inflation, and evaluating $n_{s}$ and $r$ at this field value, allows us to obtain the predictions of the model that can be contrasted to the Planck collaboration results [6]. 

\section{Analysis and results}

Before examining the parameter space spanned by $\alpha$, $\beta$ and $f$, we consider the regime $\alpha \ll 1$ and $\beta \gg 1$. Under this assumption, we find the slow-roll parameter ${\epsilon}_{0}$ (9) to be approximately:

\begin{equation}
{\epsilon}_{0} \approx \frac{{V'}^{2}}{4\tilde{\beta}{V}^{4}}
\end{equation}

This approximation will be valid while

\begin{equation}
1+\cos{\frac{\phi}{f}} \gg \frac{1}{\sqrt{2\beta}}.
\end{equation}

Performing a change of variables $u \equiv \phi / f$, ${\epsilon}_{0}$ reads:
\begin{equation}
{\epsilon}_{0} \approx \frac{1}{4{f}^{2}\tilde{\beta}}\frac{{(\frac{d}{du}V)}^{2}}{{V}^{4}}
\end{equation}
where we have factored out the dependence of $f$ given the form of $V$ considered in Natural Inflation. Under the following rescalings:
\begin{equation}
f \rightarrow Cf, \qquad \beta \rightarrow {C}^{-2}\beta
\end{equation}
where $C$ is an arbitrary constant, it is explicit that ${\epsilon}_{0}$ remains invariant.

Consider now the number of e-folds $N$ given by (14) and the expression (11). Under the approximation $\alpha \ll 1$ and $\beta \gg 1$, taking into account the range of validity (16), we get:
\begin{equation}
N \approx \int_{{\phi}_{0}}^{{\phi}_{e}}-\frac{2\tilde{\beta}{V}^{3}}{V'} d{\phi}
\end{equation}

Suppose that, given ${\phi}_{e}$, the value of the field ${\phi}_{0}$ which gives a desired number of e-folds $N$ is known. Using again the definition (11) and under $f \rightarrow Cf$ and $\beta \rightarrow {C}^{-2}\beta$:

\begin{equation}
\tilde{N} \approx \int_{{\phi}_{0}'}^{{\phi}_{e}'}{-2\tilde{\beta}{C}^{-2} \frac{{V\big(\frac{\phi}{Cf}\big)}^{3}}{ \frac{d}{d\phi}V\big(\frac{\phi}{Cf}\big)} d\phi} =  \int_{{\phi}_{0}'/C}^{{\phi}_{e}'/C}{-2\tilde{\beta} \frac{{V\big(\frac{\tilde{\phi}}{f}\big)}^{3}}{ \frac{d}{d\tilde{\phi}}V\big(\frac{\tilde{\phi}}{f}\big)} d\tilde{\phi}}
\end{equation}

\vspace{2mm}

where we defined $\tilde{\phi} \equiv \phi/C$. Demanding $\tilde{N} = N$ and taking into account that both integrands are equal, we find that ${\phi}_{e}' = C{\phi}_{e}$ and ${\phi}_{0}' = C{\phi}_{0}$. That is, the values of $\phi$ giving the desired number of e-folds $N$ transform as  ${\phi}_{e} \rightarrow C{\phi}_{e}$, ${\phi}_{0} \rightarrow C{\phi}_{0}$  under $f \rightarrow Cf$ and $\beta \rightarrow {C}^{-2}\beta$.

All the quantities defined in (9) and (10), approximated using (11) and under the conditions $\alpha \ll 1$, $\beta \gg 1$ and (16), are also invariant under the rescalings (18):

\begin{equation}
{\epsilon}_{1}=\frac{{\epsilon}_{0}'}{{\epsilon}_{0}}\frac{\dot{\phi}}{H} \approx -\frac{1}{2{f}^2\tilde{\beta}} \frac{ \frac{d}{du}{\epsilon}_{0}  \frac{d}{du}V }{{\epsilon}_{0} {V}^{3}}
\end{equation}
\begin{equation}
{l}_{0}=\frac{F'}{F}\frac{\dot{\phi}}{H} \approx -\frac{1}{2{f}^2\tilde{\beta}} \frac{ \frac{d}{du}F  \frac{d}{du}V }{F {V}^{3}}
\end{equation}
\begin{equation}
{l}_{1}=\frac{{l}_{0}'}{{l}_{0}}\frac{\dot{\phi}}{H} \approx -\frac{1}{2{f}^2\tilde{\beta}} \frac{ \frac{d}{du}{l}_{0} \frac{d}{du}V }{{l}_{0} {V}^{3}}.
\end{equation}

This means that $n_{s}$ (12) and $r$ (13) are not changed under $f \rightarrow Cf$, $\beta \rightarrow {C}^{-2}\beta$ which, in turn, will be evaluated at the invariant value ${u}_{0}={\phi}_{0}/f = C{\phi}_{0}/Cf$. Therefore, we can conclude that after finding a pair of parameters $f$ and $\beta$ that give $n_{s}$ and $r$ in agreement with the results of the Planck collaboration [6], the model with parameters $Cf$ and $C^{-2}\beta$ will also agree with them.

\vspace{3mm}

By inspecting numerically the parameter space using the full expressions (9), (10) and (11) with $\alpha \ll 1$ and in the sub-Planckian regime $f \ll 1$, we find that for $f=0.01$ and $\beta$ in the range $3.5 \times 10^{4} - 2.5 \times 10^{5}$ the predicted $n_{s}$ and ${r}$ are in agreement with the Planck collaboration results, as shown in Fig. 1, meaning that they are found within the 95\%  and 68\% confidence level (CL) contours. We find that $\epsilon_{1}$ is the first slow-roll parameter to become approximately equal to 1, being the others less than unity, and therefore defining the end of slow-roll inflation (i.e. $\epsilon_{1}(\phi_e) \simeq 1$). For this range of $\beta$ we find ${\phi}_{e}/f$ to have values from 1.72 to 2.13. This implies that condition (16) will be fulfilled all the way down from $\phi_{0}/f$ to ${\phi}_{e}/f$. Using then the approximate symmetry under rescalings (18) we can arbitrarily decrease $f$ by increasing $\beta$ and, for instance, we would obtain $\beta$ in the range $3.5 \times 10^{6} - 2.5 \times 10^{7}$ for $f={10}^{-3}$, and $3.5 \times 10^{8} - 2.5 \times 10^{9}$ for $f={10}^{-4}$. In Fig. 1 the results for $\alpha = 0.1$ are also shown, in which case the approximation is not very accurate, but we find that even in this scenario we obtain predictions falling within the contours of Planck data.

\vspace{3mm}

\begin{figure}
\begin{center}
\vspace{3mm}
\includegraphics[scale=0.45]{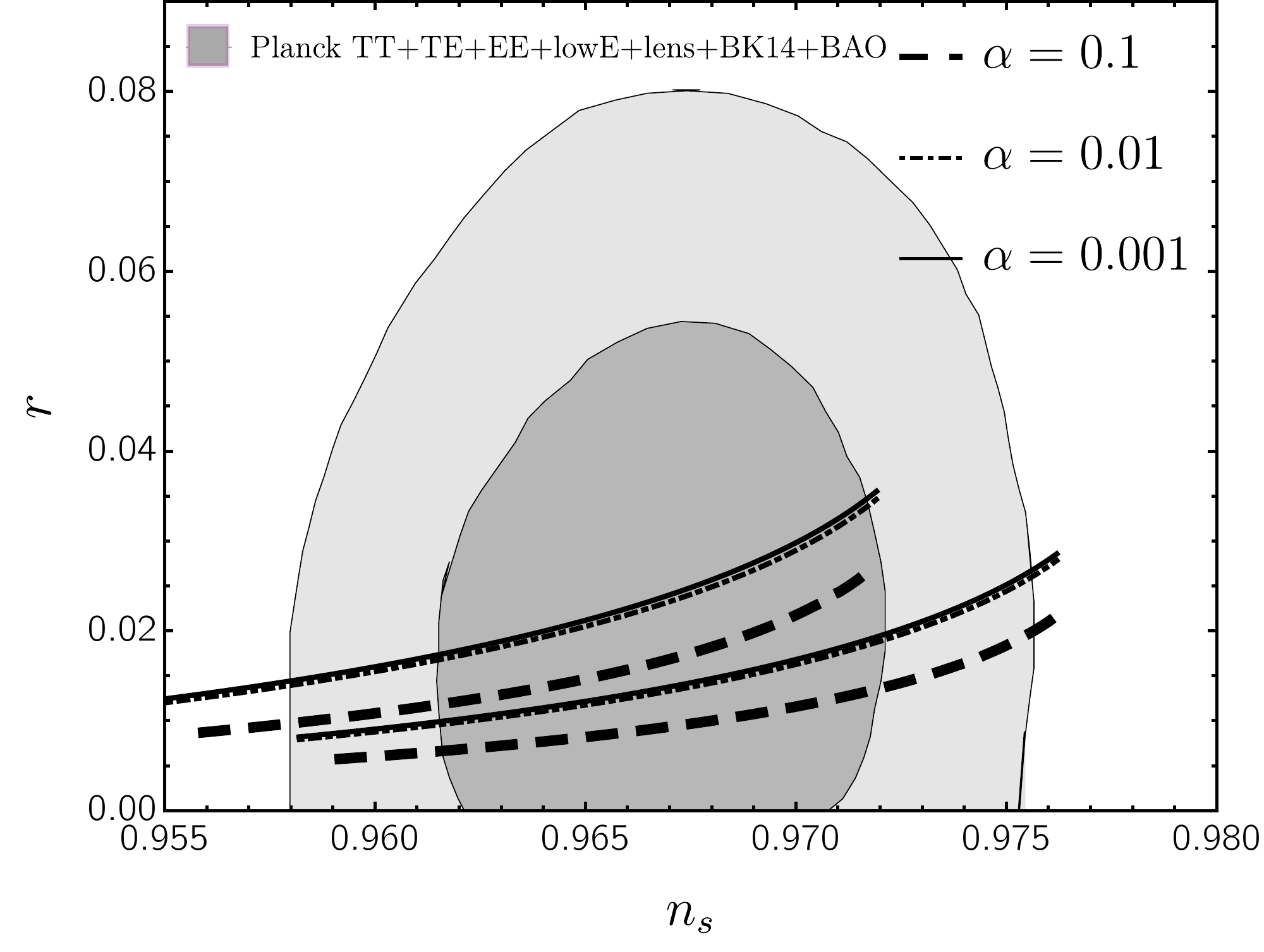}
\caption {Results obtained by running $\beta$ from $3.5\times10^{4}$ to $2.5\times10^{5}$ (from left to right) for several values of $\alpha$ and $f=0.01$, taking $N=50$ (upper lines) and $N=60$ (lower lines). The cases of $\alpha = 0.01$ and $\alpha=0.001$ are almost indistinguishable.  Light and dark shading correspond, respectively, to 95\% C.L. and 68\% C.L. of Planck collaboration constraints.}
\end{center}
\end{figure}

Finally, we constrain the value of the scale $\Lambda$ by imposing the normalization of the power spectrum [15] given by the Planck collaboration [6]

\begin{equation}
P_{\xi }\sim \frac{H^2}{8\pi ^2}\frac{1}{\epsilon _0}= \frac{V}{24{\pi} ^{2}F}\frac{1}{\epsilon _0}  \simeq 2.1 \times 10^{-9}
\end{equation} 

giving $\Lambda \sim 3 \times 10^{-3} - 5\times 10^{-3}$ for $f$ and the range of $\beta$ considered before.

\section{Conclusions}

We have shown that the embedding of Natural Inflation in a more general scalar-tensor theory, with couplings respecting the periodicity and the shift symmetry of the original model, can drive slow-roll inflation for $f \ll 1$ in agreement with Planck collaboration results. For example, for $f=0.01$ and $\beta$ in the range $3 \times 10^{4} - 1.5 \times 10^{5}$ we obtain spectral indices $n_s$ and tensor-to-scalar ratios $r$ within the 95\% and 68\% CL contours of Planck data, giving $\Lambda \sim 10^{-3}$. 

Furthermore, in the limit of small coupling to the Ricci scalar, i.e. $\alpha \ll 1$, given that $\beta \gg 1$ for results in agreement with Planck, the theory presents a symmetry under the rescalings $f \rightarrow Cf$ and $\beta \rightarrow C^{-2}\beta$, leaving the values of $n_s$ and $r$ invariant. This remarkable property allows to arbitrarily decrease the value of the periodicity scale $f$ by increasing the coupling to the kinetic term, solving the issue of Natural Inflation super-Planckian periodicity scales $f$ and alleviating the current tension of Natural Inflation with Planck data.

Finally, we are aware of the fact that the UV completion of the model has not been studied in the present paper, but the analysis of this issue is beyond the scope of this letter. Therefore, this issue is expected to be addressed in future work.

\begin{acknowledgments}
I would like to thank Alessio Notari and Ricardo Z. Ferreira for introducing me to Natural Inflation.
\end{acknowledgments}

\begin{center}
\noindent\rule{8cm}{0.4pt}
\end{center}

{\small
[1] K. Freese, J. A. Frieman, and A. V. Olinto, Phys. Rev. Lett. \textbf{65}, 3233 (1990).

[2] T. Banks, M. Dine, P. J. Fox, and E. Gorbatov, JCAP \textbf{0306}, 001 (2003), hep-th/0303252.

[3] T. Rudelius, JCAP \textbf{1509}, 020 (2015), 1503.00795.

[4] T. Rudelius, JCAP \textbf{1504}, 049 (2015), 1409.5793.

[5] M. Montero, A. M. Uranga, and I. Valenzuela, JHEP \textbf{08}, 032 (2015), 1503.03886.

[6] N. Aghanim et al. (Planck collaboration), arXiv:1807.06209.

[7] N. Arkani-Hamed, H.-C. Cheng, P. Creminelli, and L. Randall, Phys. Rev. Lett. \textbf{90}, 221302 (2003), hep-th/0301218.

[8] J. E. Kim, H. P. Nilles, and M. Peloso, JCAP \textbf{0501}, 005 (2005), hep-ph/0409138.

[9] S. Dimopoulos, S. Kachru, J. McGreevy, and J. G. Wacker, JCAP \textbf{0808}, 003 (2008), hep-th/0507205.

[10] M. Gerbino, K. Freese, S. Vagnozzi, M. Lattanzi, O. Mena, E. Giusarma, and S. Ho, Phys. Rev. D \textbf{95} (2017), no.4, 043512, 1610.08830.

[11] R. Z. Ferreira and A. Notari, Phys. Rev \textbf{D97}, 063528 (2018), 1711.07483.

[12] M. P. Hertzberg, JHEP \textbf{11}, 023 (2010), 1002.2995.

[13] F. L. Bezrukov and M. Shaposhnikov, Phys. Lett. B \textbf{659}, 703 (2008), 0710.3755.

[14] S. Capozziello, G. Lambiase, Gen. Rel. Grav 31, 1005 (1999), gr-qc/9901051.

[15] S. V. Sushkov, Phys. Rev. D \textbf{80}, 103505 (2009), 0910.0980.

[16] L. N. Granda, JCAP \textbf{07}, 006 (2010), 0911.3702.

[17] L. N. Granda, W. Cardona, JCAP \textbf{07}, 021 (2010), 1005.2716.

[18] S. Capozziello, G. Lambiase and H. J. Schmidt, Annalen Phys. \textbf{9}, 39 (2000), gr-qc/9906051.

[19] C. Germani and A. Kehagias, Phys. Rev. Lett \textbf{105}, 011302 (2010), 1003.2635.

[20] R. Z. Ferreira, A. Notari, and G. Simeon, JCAP \textbf{2018}, 021 (2018), 1806.05511.

[21] S. Aashish and S. Panda, JCAP \textbf{06}, 009 (2020), 2001.07350.

[22] C. Germani, and A. Kehagias, Phys. Rev. Lett. \textbf{106}, 161302 (2011), 1012.0853.

[23] L. N. Granda and D. F. Jimenez, JCAP \textbf{09}, 007 (2019), 1905.08349.

\end{document}